\documentclass[cits]{JINST}
\usepackage{amsmath}
\usepackage{fnpos}
\makeFNbottom
\title{A new method based on noise counting
to monitor the frontend electronics of the LHCb muon detector}

\author{
L.~Anderlini$\,^a$,
R.~Antunes Nobrega$\,^b$\thanks{Now at Universidade Federal de Juiz 
de Fora, Juiz de Fora, Brazil.}~,
W.~Bonivento$\,^c$,
L.~Gruber$\,^d$\thanks{Now at Technische Universit\"at Wien, 
Austria.}\,,
A.~Kashchuk$\,^e$,
O.~Levitskaya$\,^e$,~
O.~Maev$\,^e$,~
G.~Martellotti$\,^b$,~ 
G.~Penso$\,^{b,f}$,~
D.~Pinci$\,^b$\thanks{Corresponding author.}~,~
\hbox{A. Sarti$\,^{f,\,g}$}, ~~~~~~~~~
B. Schmidt$\,^d$
\\
\llap{$^a$} Sezione INFN di Firenze, Firenze, Italy \hfill \\
\llap{$^b$} Sezione INFN di Roma, Roma, Italy \hfill \\
\llap{$^c$} Sezione INFN di Cagliari, Cagliari, Italy \hfill \\
\llap{$^d$} European Organisation for Nuclear Research (CERN), Geneva, 
Switzerland \hfill \\
\llap{$^e$} Petersburg Nuclear Physics Institute, Gatchina, 
St-Petersburg, Russia \hfill \\
\llap{$^f$} Sapienza, Universit\`a di Roma, Roma, Italy \hfill \\
\llap{$^g$} Laboratori Nazionali di Frascati dell'INFN, Frascati, 
Italy 
\hfill \\
E-mail: \email{davide.pinci@roma1.infn.it}
}

\abstract{A new method has been developed to check the correct behaviour 
of the frontend electronics of the LHCb muon detector. 
This method is based on the measurement of the electronic noise
rate at different thresholds of the front-end discriminator. 
The method was used to choose the optimal discriminator thresholds.
A procedure based on this method was implemented in the detector 
control system and allowed the detection of a small percentage of
front-end channels which had deteriorated.
A Monte Carlo simulation has been performed to check 
the validity of the method.
}
\keywords{Muon spectrometers; Front-end electronics for detector readout;
electronic noise}

\begin{document}

\section{Introduction}
The muon detector of the LHCb experiment \cite{jinst,muonpaper} 
is composed of 5 stations \hbox{(M1$-$M5)} placed along 
the beam axis and with a total area of 435~m$^2$.
It identifies muons with an efficiency higher than $99~\%$ 
per station \cite{muonpaper},
provides fast information for the high-$p_{\mathrm T}$ 
muon trigger at the earliest 
level \hbox{(Level-0)} and muon identification for the high-level 
trigger and offline analysis.
Each station is divided in 4 regions \hbox{(R1$-$R4)}, 
with increasing distance from the beam pipe.
The linear dimensions of the regions R1, R2, R3, R4, and their 
segmentations scale in the ratio 1:2:4:8. With this geometry,
the particle flux and channel occupancy are expected to be roughly the 
same over the four regions of a given station.
Multi-wire proportional chambers (MWPC) are used everywhere,
except in the inner region of station M1 (M1R1) where 
the expected particle rate exceeds safety limits for MWPC ageing. 
In this region \hbox{triple-GEM} detectors 
\cite{jinst} are used.
The detector comprises 1380 chambers with 122112 readout channels.
The front-end electronics of each channel are equipped with the 
CARIOCA chip \cite{carioca} which performs the
signal amplification, shaping and discrimination.
The detector is designed to work for many years in a high 
radiation environment. Therefore
to minimize the ageing effects, the chambers should 
operate at the lowest possible 
gain \cite{gain} compatible with the high efficiency required \cite{burkhard}, 
and the discriminator thresholds should be as 
low as possible compatible with an acceptable noise rate.

During the many years forseen for the data taking of the experiment, the 
stability of the readout channels must be monitored and any
possible ageing effect or breakdown of the front-end electronics 
should be quickly and effectually detected. In this paper we 
describe a new method \cite{anatoli, anatoli2} used to keep
all the readout channels of the MWPCs of the LHCb muon 
detector under control. 
This method consists \cite{lukas} in counting
the electronic noise rate as a function of the
threshold of the discriminator\footnote{Hereafter this procedure is 
briefly called "threshold scan".} for each channel. 
It allows to check the correct
behaviour of the frontend electronics and to choose the optimal
values of the discriminator thresholds.

The threshold scan procedure was implemented in the LHCb detector control 
system and is systematically applied during the operation of 
the experiment.
\vspace*{-1truemm}
\section{The front-end electronics}
\label{section1}
To meet the required condition on rate capability and resolution,
the MWPCs of the muon detector were partitioned in pads of different 
size and capacitance. The readout was performed on the cathodes 
or/and on the anodes of the chambers. 
Each channel of the front-end electronics comprises an amplifier,
a shaper and a discriminator. 
Each CARIOCA chip comprises 8 readout channels and two CARIOCAs were 
mounted on a front-end board which therefore performs
the readout of 16 chamber pads.  

\subsection{The amplifier}
\begin{figure}[h]
\vspace*{3mm}
\begin{center}
\includegraphics[width=.8\textwidth]{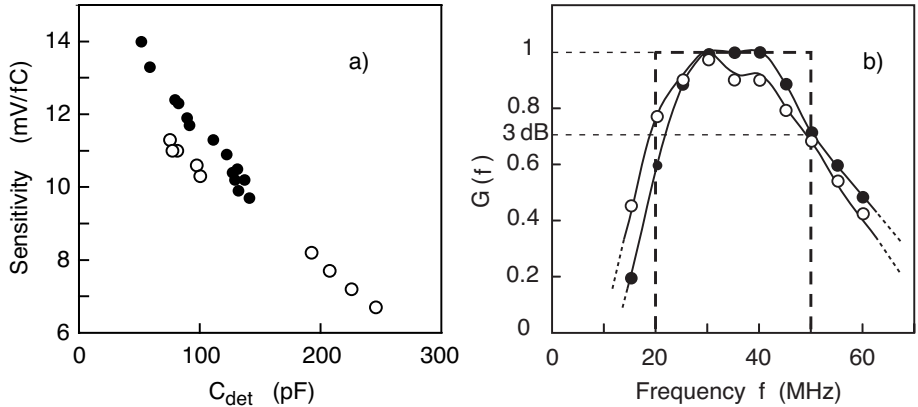}
\end{center}
\vspace*{-5truemm}
\caption {
a): Measured sensitivity of the amplifier.
b): Measured gain $G$ of the amplifer for a sinusoidal 
signal as function of frequency~$f$. The curves are 
normalized to 
their maximum value.
The rectangular approximation to the frequency spectrum of the
amplifier gain, as used in the MC
simulation, is indicated with the 3dB limits at 20 and 50 MHz.
The lines are to guide the eye.
In both figures the open (full) points
refer to negative (positive) input polarity.
}
\label{amplifier}
\end{figure}

The amplifier of the CARIOCA chip can be wired to handle 
either negative polarity for anode channels or positive
polarity for cathode pads.
The sensitivity of the amplifier, defined as the output pulse amplitude
per unit $\delta$-like charge, was
measured as a function of the capacitance
($C_{det}$) of the detector. In figure~\ref{amplifier}a the dependence
of the sensitivity on the pad capacitance is shown.

The gain of the amplifier (figure \ref{amplifier}b) was measured as 
function of frequency at $C_{det} \simeq 0$. The 3dB limits were found to 
lie 
near 20 and 50MHz, for positive and negative input polarity.
The output of each amplifier, which has the same sign for cathode 
and anode readout, is sent to a discriminator.

\subsection{The discriminator}

\begin{figure}[h]
\vspace*{3mm}
\begin{center}
\includegraphics[width=.6\textwidth]{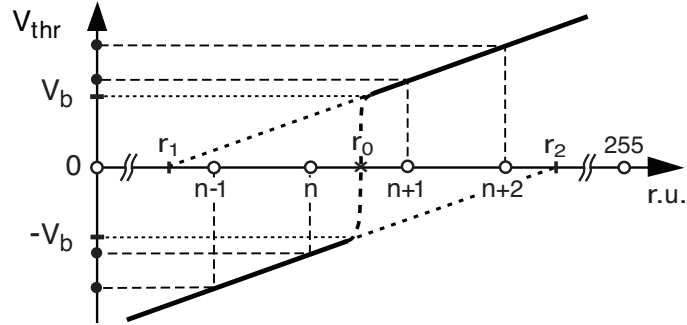}
\end{center}
\vspace*{-5truemm}
\caption {
Dependence of the threshold ($V_{thr}$) of the CARIOCA discriminator
around zero on the DAC register units ($r.u.$). 
}
\label{discriminator1}
\end{figure}

A 8-bit digital-to-analog converter (DAC) with 256 steps 
allows to change the threshold of each discriminator
in steps of 2.35~mV which corresponds to a single
DAC register unit ($r.u.$). The threshold can vary from positive to
negative values, with the exception of a ``bias'' interval
\hbox{$V_b \simeq \pm 26$~mV} (figure~\ref{discriminator1}) around zero.
In this region the transition between positive and negative threshold 
occurs within one DAC register.
The amplitude of the signal at the input of the amplifier 
depends on the capacitance of the pad read by the front-end.

\begin{figure}[h]
\vspace*{3mm}
\begin{center}
\includegraphics[width=.85\textwidth]{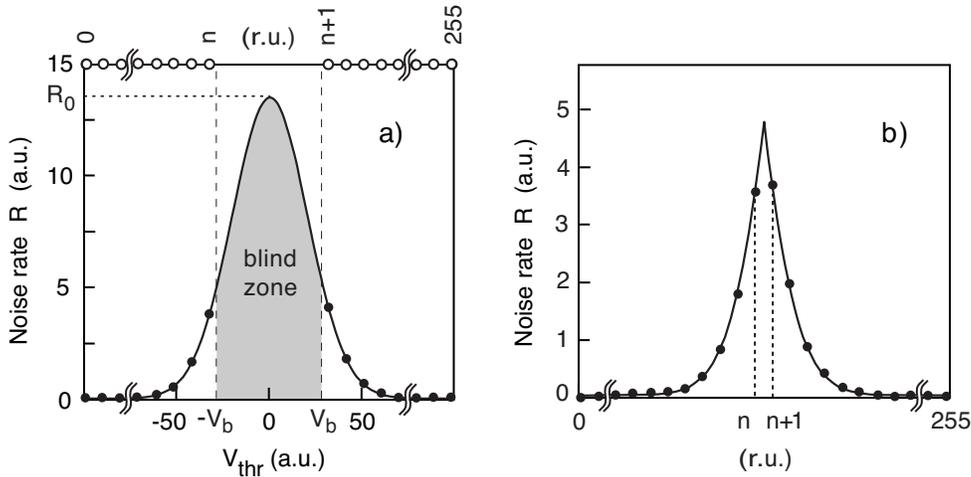}
\end{center}
\vspace*{-5truemm}
\caption {
a): Noise counting rate of an ideal discriminator as a 
function of its threshold (continuous curve). 
With the CARIOCA discriminator the threshold values 
comprised in the shaded area cannot be set and the rate at
zero threshold ($R_0$) cannot be measured.
In the upper scale the 
DAC registers which set the thresholds are schematically reported. 
b): Typical experimental curve representing the noise counting rate of 
a CARIOCA discriminator as a function of the DAC registers.
}
\label{discriminator2}
\end{figure}

For an ideal discriminator with no bias interval, the dependence of 
the noise counting rate on the threshold is  a 
bell-shaped distribution (figure~\ref{discriminator2}a). 
Two parameters of this distribution will be considered
in the following: 
the counting rate at zero threshold ($R_0$) and the r.m.s.~($\sigma_R$) of 
the distribution.

For the CARIOCA discriminator, because of its bias interval, only 
the tails of the bell-shaped distribution can be measured. 
A threshold scan will therefore result in a 
double-arm curve with a cusp 
(figure~\ref{discriminator2}b), the right (left) 
arm corresponding to the positive (negative) tails.

To monitor the MWPC readout channels a check of the 
stability of the $\sigma_R$ values of all the 122k channels
is periodically performed. 
For each channel, $\sigma_R$ must be inferred from the corresponding 
threshold scan distribution. This evaluation would be 
more precise if the shape of the distribution were known. 
To find out this distribution in the present 
experimental situation, a Monte Carlo (MC) simulation was performed.
 
\section{The Monte Carlo simulation}

The MC simulation comprises the generation of the noise signal
sent to the discriminator by the amplifier and the calculation of the
output rate of the discriminator as a function of its threshold.
The effect of the dead time of the CARIOCA was evaluated.
The bias interval was not considered in the MC but
will be taken into account later.

\subsection{The noise signal}
The noise at the input of the front-end electronics is assumed to 
have a flat frequency spectrum\footnote{The detector capacitance may
modify the noise frequency spectrum. This effect has negligible
consequence on the experimental results (see section \ref{expres1}) and 
therefore was neglected in the MC.} (white noise), resulting
from the superposition of a great number of contributions which occur 
randomly. This noise is then filtered by the amplifier and sent to 
the discriminator. Therefore the passband of the amplifier 
determines the frequency spectrum of the noise at the discriminator input. 
In the MC this passband was approximated to a rectangular one,
extending from 20~MHz to 50~MHz, which corresponds to the 3~dB 
bandwith of the amplifier (figure~\ref{amplifier}b).
This approximation will allow
to compare easily the MC results with Rice's theory \cite{rice}, 
and is expected to give a satisfactory representation of the 
experimental situation.

The time dependence of the noise signal $V(t)$ at the input of the 
discriminator was generated as a superposition of 
\hbox{$N = 60000$} sinusoids according to the formula:\hfill
\vskip -6pt
\begin{equation} 
\label{sum}
V(t) = A \sum \limits_{k=1}^N sin(2\pi f_k t +\phi_k)
\end{equation}
\vskip 8pt
\noindent
The frequency $f_k$ and the phase $\phi_k$ were randomly extracted
in the intervals $20-50$~MHz and $0-2\pi$ respectively.
The noise signal was calculated every 3~ns, assuming $A=1$ in 
equation~\ref{sum}. 
With an upper frequency of 
$\sim 50$~MHz this sampling of the noise is 
sufficiently dense to give a correct description of the signal.

\begin{figure}[t]
\rule{10mm}{0pt}
\vspace*{3mm}
\begin{center}
\includegraphics[width=.95\textwidth]{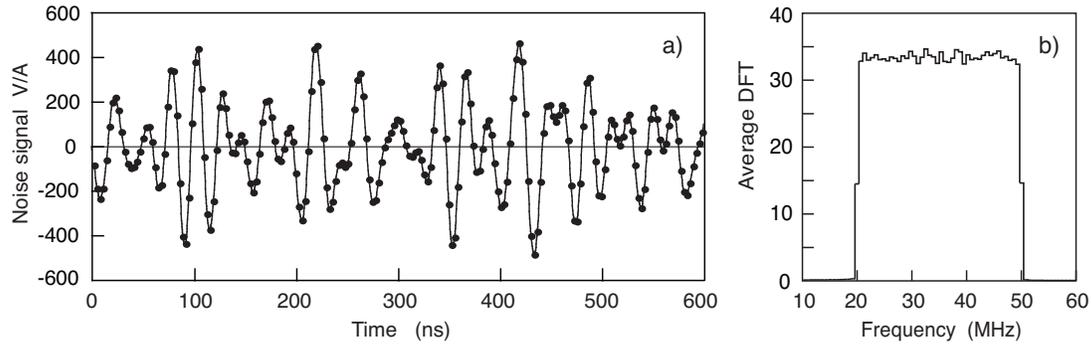}
\end{center}
\vspace*{-4truemm}
\caption {
a): A typical simulated noise signal. The points are the values 
calculated every 3 ns. b): Discrete Fourier transform of 
10$^6$ consecutive noise values, corresponding to a time interval of 3~ms. 
The DFT values have been averaged on 0.7 MHz bins.
}
\vskip 1 truemm
\label{signal}
\end{figure}
 
The MC generates $2\times 10^8$ consecutive values of the noise signal, 
corresponding to a time interval of 600 ms. 
As shown in the next section, this time interval gives a number of 
threshold crossings sufficiently high to simulate with good statistics 
the tails of the threshold scan distribution. 
In figure~\ref{signal}a we report a small part of the simulated noise 
signal. 

To check the accuracy of the noise simulation 
a discrete Fourier transform (DFT) of $10^6$ consecutive noise values 
(corresponding to a  3 ms time interval) was performed. 
The results, shown in 
figure~\ref{signal}b, reproduce quite well the bandwidth assumed in the 
MC, 
giving confidence in the signal generation procedure.

\begin{figure}[b]
\begin{center}
\includegraphics[width=.9\textwidth]{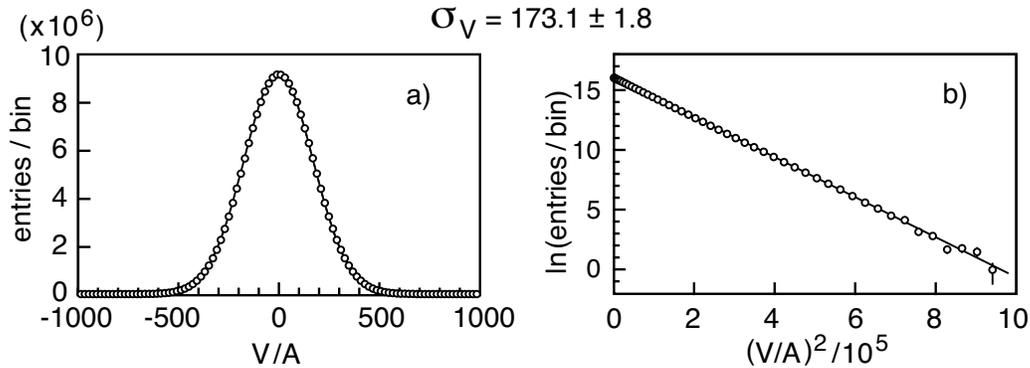}
\end{center}
\vspace*{-4truemm}
\caption {
a): Distribution of the $2 \times 10^8$ values of the noise 
amplitude $V/A$ (equation~\protect \ref{sum}). 
In figure b) the same distribution is 
reported with logarithmic-quadratic scales.
}
\label{sigmaV}
\end{figure}

The distribution of the noise amplitudes, calculated with 
equation \ref{sum} 
every 3~ns, is reported in figure~\ref{sigmaV}a. The 
r.m.s.~of this 
distribution, deduced from a Gaussian fit, is $\sigma_V = 173.1 \pm 
1.8$. 
The same points, reported in figure~\ref{sigmaV}b with 
logarithmic-quadratic scales, appear to be perfectly aligned up to
$(V/A)^2 \simeq 9 \times 10^5$. This confirms that the 
distribution of the noise amplitudes is, with 
an excellent approximation, a Gaussian up to $\sim 6\sigma$.
As a further check $\sigma_V$ was also calculated from the series of 
generated 
values of $V(t)$, according to the formula:
\vspace*{7truemm}
\begin{equation} 
\label{average}
\sigma_V =\left\{ \overline{(V(t)/A)^2} \right\} ^{1/2} = 
\left\{ 
\overline{ 
\left[ \sum \limits_{k=1}^N sin(2\pi f_k t +\phi_k) \right] ^2
}  
\right\} ^{1/2} 
= \sqrt{N/2} = 173.2
\vspace*{7truemm}
\end{equation}
\noindent
\noindent a value in quite good agreement with the one obtained with the 
Gaussian fit. 

The quantity $\sigma_V$ is related to the equivalent noise 
charge (ENC) which is defined as the $\delta$-like charge sent to the 
front-end which delivers an output signal equal to $\sigma_V$.

\vspace*{6truemm}
\subsection{The discriminator counting rate}
\vspace*{6truemm}

\begin{figure}[b]
\vspace*{3mm}
\begin{center}
\includegraphics[width=.9\textwidth]{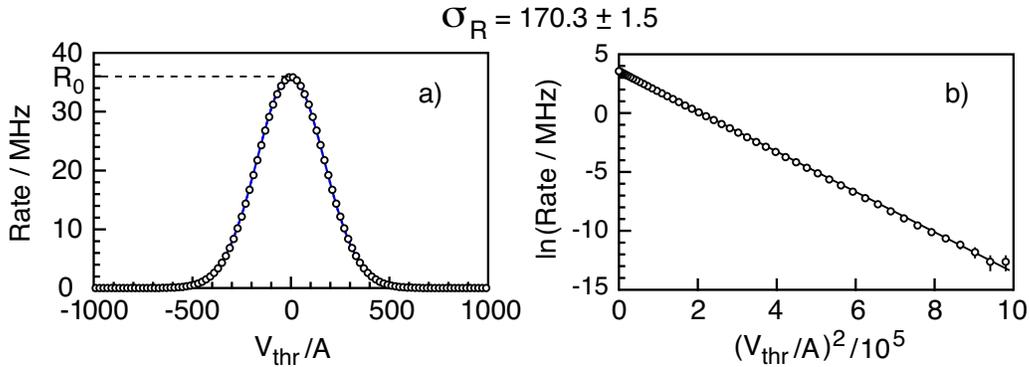}
\end{center}
\vspace*{-8truemm}
\caption {
a) Counting rate of the discriminator as a function of its
threshold. b) Same as in a) but presented with logarithmic-quadratic
scales.
}
\vspace*{1truemm}
\label{sigmaR}
\end{figure}

To identify the threshold crossings, a continuous noise signal 
was necessary. This was obtained by a linear interpolation
between consecutive pairs of MC generated points.
The discriminator was considered to fire when this
signal crosses the threshold with positive (negative) slope for positive 
(negative) thresholds. The counting rate ($R$) of
the discriminator was then determined as a function of its 
threshold\footnote{In this section the discriminator was assumed 
to have no bias interval.}.
The result is reported in figure~\ref{sigmaR}a.
The same points, reported in figure~\ref{sigmaR}b with
logarithmic-quadratic scales, are perfectly aligned up to
$(V_{thr}/A)^2 \simeq 9 \times 10^5$. This confirms 
\hbox{\cite{rice,kac,ivanov}}
that the distribution of the noise counting rate (figure~\ref{sigmaR}a) 
is, with
an excellent approximation, a Gaussian up to $\sim 6\sigma$.
A Gaussian fit to this distribution gives an r.m.s. value 
\hbox{$\sigma_R = 170.3 \pm 1.5$}.

Within the errors, the MC results shown in figure~\ref{sigmaV} 
and figure~\ref{sigmaR} confirm the relation predicted by Rice 
\cite{rice}:
\vspace*{3truemm}
\begin{equation}
\label{sigmaVR}
\sigma_V = \sigma_R \equiv \sigma
\end{equation}
\vspace*{1truemm}

\noindent
which links the r.m.s. of two a priori 
different distributions, and allows to evaluate $\sigma_V$ and therefore 
the 
ENC of a channel, by measuring its noise rate as a function of 
the discriminator threshold.

With an ideal discriminator without any bias interval, the counting rate
would depend on $V_{thr}$ (figure~\ref{sigmaR}a) according to the formula:
\vspace*{5truemm}
\begin{equation}
\label{gaussian}
R = R_0 \; \mathrm{exp}\;[-V_{thr}^2/(2 \sigma^2)]
\end{equation}
\vspace*{4truemm}

\noindent
The counting rate ($R_0$) at zero threshold, calculated 
from a fit to the MC results (figure~\ref{sigmaR}), is 
$35.7 \pm 1.7$~MHz, in quite good agreement with 
the value 
calculated with the 
Rice formula\footnote{The Rice formula \cite{rice} differs by a factor 
2 from equation~\ref{rzero} because in that paper both the positive-slope 
and negative-slope threshold crossing are counted.} for a flat 
bandwidth: \hfill \\
\vspace*{-1truemm}

\begin{equation} 
R_0 = \left[ \frac {f_b^3 - f_a^3} {3(f_b - f_a)} \right] ^{1/2} = 36.1 \; 
\mathrm{MHz} 
\label{rzero}
\end{equation}
\vspace*{9truemm}  

\noindent   
where $f_a = 20$~MHz and $f_b = 50$~MHz are the lower and upper cut-off 
frequencies of the bandwidth.

\begin{figure}[b]
\vspace*{5mm}
\begin{center}
\includegraphics[width=.9\textwidth]{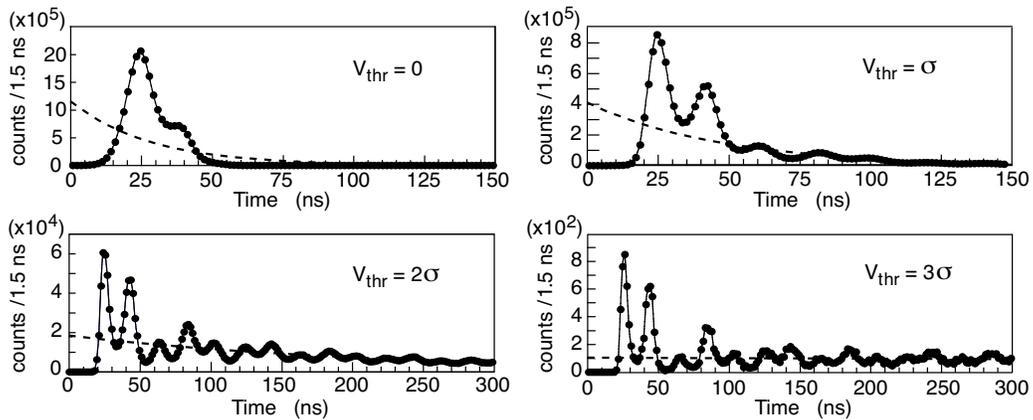}
\end{center}
\vspace*{-5truemm}
\caption {Monte Carlo predictions for the distribution
of the time intervals between
two consecutive threshold crossing, for four different threshold
values. The dashed curves represent the
same distributions if the crossing times were completely uncorrelated.
}
\label{crosstime}
\end{figure}

If the exact passband $G(f)$ (figure~\ref{amplifier}b) of the front-end 
amplifier is considered,
the rate at zero threshold may be calculated 
numerically from the general Rice formula \cite{rice}:
\vspace*{10 truemm}
\begin{equation}
R_0 = \left[ \frac{\int\limits_0^{\infty} \; f^2 \: W(f) \;
\mathrm{d}f}{\int\limits_0^{\infty} \; W(f) \; \mathrm{d}f}
\right] ^{1/2} =
\left[ \frac{\int\limits_0^{\infty} \; f^2 \: G^2(f) \;
\mathrm{d}f}{\int\limits_0^{\infty} \; G^2(f) \; \mathrm{d}f}
\right] ^{1/2}
\label{R_02}
\end{equation}
\vspace*{7 truemm}

\noindent 
where $W(f)$ is the power spectrum of the noise at the input of the 
discriminator and $W(f) \propto G^2(f)$ for white noise.
From equation \ref{R_02} it
turns out to be $R_0 \simeq 38 \pm 3$~MHz
for negative polarity and of $R_0 \simeq 40 \pm 3$~MHz for positive 
polarity, the error being due to non-perfect knowledge of $G(f)$.

At low threshold, and therefore at high counting rate, the effect of 
the dead time of the CARIOCA ($\sim 55$~ns for noise 
signal) must be considered. To evaluate this effect, 
the distribution of the time intervals 
between two consecutive threshold crossings was calculated with the MC.
In figure~\ref{crosstime} this distribution is reported for four 
threshold values\footnote{The oscillating behaviour of these 
distributions is in agreement with the results reported in 
ref.~\cite{longuet}.}. At the lowest thresholds reached in the threshold 
scans 
(about 2$-$2.5$ \sigma$, see section~\ref{expres1}) the loss in couting 
rate due 
to the dead time is about 15$-$20~\%. 
This effect corresponds to a variation <~2~\%  on $\sigma$  
and was therefore neglected. 

\vspace*{-2truemm}

\section{Experimental results}
\vspace*{-1truemm}

\subsection{Threshold determination}
\label{expres1}
The threshold scan of all the readout channels is performed
regularly between two data taking periods. 
For each channel and for each threshold the noise is counted  during 200~ms. 

\begin{figure}[b]
\begin{center}
\includegraphics[width=.9\textwidth]{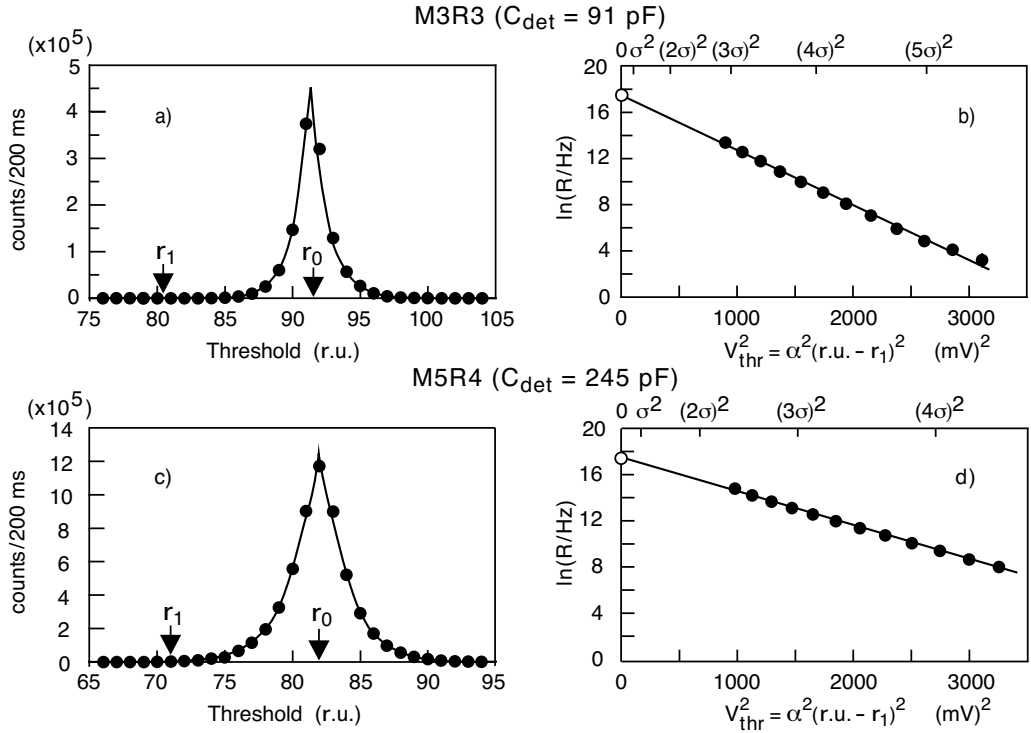}
\end{center}
\vspace*{-6truemm}
\caption {
a): Experimental threshold scan of a readout channel of region 
M3R3. In b) the right-arm of this distribution is represented with 
logarithmac-quadratics scales. The open point at zero threshold 
is the counting rate predicted by Rice theory and by the MC.  The straight 
line is the best fit to the points (see text). c) and 
d): same as a) and b) for a readout channel belonging to region M5R4.
The larger detector capacitance ($C_{det}$) results in a larger $\sigma$.
}
\label{exp_sigma}
\vspace*{0truemm}
\end{figure}

As an example, two threshold scans are shown
in figure~\ref{exp_sigma}a and \ref{exp_sigma}c, one for a readout 
channel belonging to region M3R3 (with a cathode readout and a pad
capacitance of $\sim 91$ pF) and the other 
to M5R4 (with an anode readout and a capacitance of $\sim 245$ pF). 
The two arms of the experimental distributions being almost 
symmetrical, only the right-arms are considered\footnote{The left-arms
can also be analized. In that case $r_2$ (figure~\ref{discriminator1}) 
replace $r_1$ in all the equations.} in the following.

Taking into account the characteristics of the CARIOCA discriminator, the bias
interval ($V_b$) and the threshold ($V_{thr}$)  depend on the 
DAC register (figure~\ref{discriminator1}) according to the relations:
\vspace*{3truemm}
\begin{align}
V_b =& \alpha \: (r_0 - r_1)  \label{bias} \\[8pt]
V_{thr}=& \alpha \: (r.u. - r_1) \qquad   (r.u. > r_0)  
\label{DACr1}
\end{align}
\vskip 5truemm
\noindent
where $\alpha = 2.35$~mV/$r.u.$, $r_0$ is the interpolated DAC register 
corresponding to the threshold jump from positive to negative 
values and $r_1$ is the interpolated register corresponding to the zero 
threshold for the right arm of the threshold scan 
(figure~\ref{discriminator1}).
The experimental data, 
were represented on the plane 
($\;(r.u. - r_1)^2\;,\;\mathrm{ln}\;R\;$)  and 
fitted to a straight line given by the equation:
\vspace*{7truemm}
\begin{equation}
\label{gaussian2}
\mathrm{ln}\; R = \mathrm{ln}\; R\,'_0 - \alpha^2 
(r.u. - r_1)^2/(2\,\sigma^2)
\vspace*{7truemm}
\end{equation}
%
\noindent where ln$\;R\,'_0$ and $\sigma^2$ are the parameters of the fit. 
The value of $r_1$ being a priori not known, the fit
was repeated for different values of $r_1$ until the value of 
$\;\mathrm{ln}\;R\,'_0$ obtained from the fit is as 
close as possible to the expected value of ln$\;R_0$, where $R_0$ is 
given by equation~\ref{R_02}. 

\begin{figure}[h]
\vspace*{3mm}
\begin{center}
\includegraphics[width=.9\textwidth]{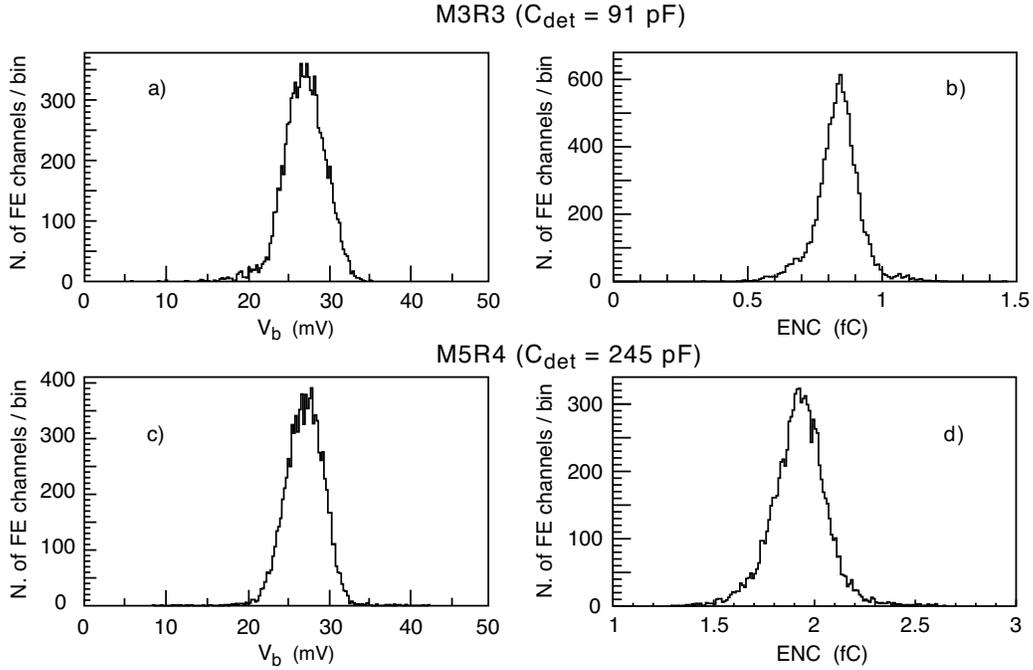}
\end{center}
\vspace*{-7truemm}
\caption {
Distribution of: a) the bias voltage; b) the ENC measured for of all the 
reaudout channels of the region M3R3.
c) and d): same as a) and b) for the readout channels belonging to region 
M5R4.
\vspace*{5truemm}
}
\label{bias_encM3R3_M5R4}
\end{figure}

\begin{figure}[t]
\vspace*{3mm}   
\begin{center}
\includegraphics[width=.9\textwidth]{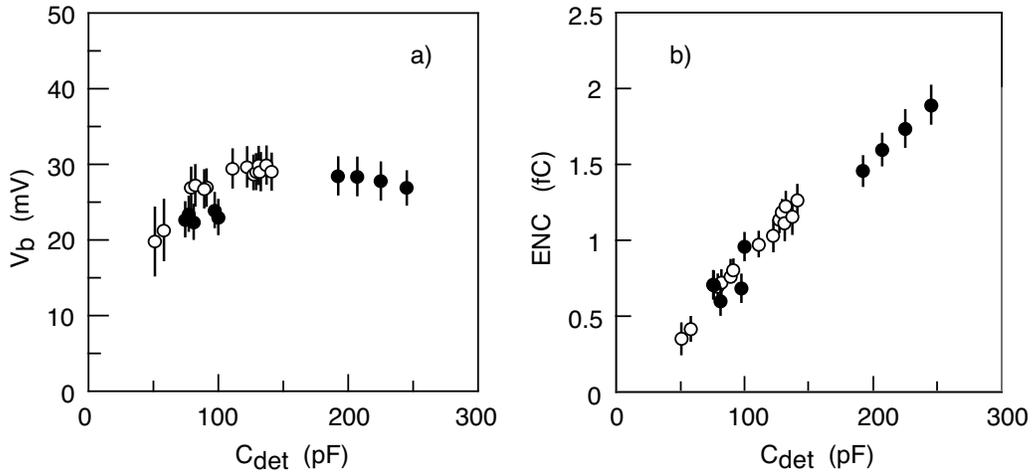}
\end{center}
\vspace*{-6truemm}
\caption {
Dependence: a) of the bias voltage ($V_b$) and b) of the 
ENC, on the detector capacitance. Open (full) points refer to
cathode (anode) readout. The error bars represent the r.m.s. of the 
distributions.
}
\label{Cdet}
\vspace*{3truemm}
\end{figure}

It is worthwhile to note that in equation~\ref{gaussian2}, $\sigma$ is 
related to $R\,'_0$ through its logarithm. Therefore the effect of the 
detector capacitance
on the noise spectrum and therefore on $R\,'_0$ has no significant 
consequence on the value of $\sigma$.

In figure~\ref{exp_sigma}b and \ref{exp_sigma}d the experimental points
belonging to the right-arms of the threshold scans are reported with
logarithmic-quadratic scales together with the results of the fitting
procedure. The agreement is quite good up to $\sim 5\, \sigma$.

From equation~\ref{gaussian2} it turns out that
the slope of the fitting line is inversely proportional 
to $\sigma^2 \;$ i.e. to $\mathrm{ENC}^2$, 
For the two readout channels considered in figure~\ref{exp_sigma} 
the results of the fits were for M3R3:
$\sigma = 4.36$~$r.u.$ $= 10.2$~mV;  
$r_0 = 91.4$~$r.u.$; $r_1 = 80.3$~$r.u.$  
and for M5R4: 
$\sigma = 5.56$~$r.u.$ $= 13.1$~mV;              
$r_0 = 82.0$~$r.u.$; $r_1 = 70.8$~$r.u.$
Taking into account the detector capacitance 
and the corresponding sensitivity 
(figure~\ref{amplifier}a and Table~\ref{setting}), the values of $\sigma$ 
correspond to an  
ENC = 0.80~fC for M3R3 and 1.89~fC for M5R4. 
In figure~\ref{bias_encM3R3_M5R4} the distribution 
of the bias voltage $V_b$ and of
the ENC are reported for all the channels of 
the regions M3R3 and M5R4.

In figure~\ref{Cdet} the mean value of the bias ($V_b$) and of the 
ENC is reported
as a function of the pad capacitance of all the different
chamber types of the muon detector. As expected, $V_b$ 
is characteristic
of the discriminator and is roughly independent of $C_{det}$, while
the increase of ENC with $C_{det}$ is approximately linear 
\cite{anghinolfi}.
\vspace*{3truemm}

A summary of the measured characteristics of the readout channels 
of different regions is reported in Table~\ref{setting}. 
The measured ENC values allow 
to set the working condition of each type of muon chamber
for the LHCb data taking.
In fact the thresholds should be sufficiently high to limit
the counting rate due to electronic noise and sufficiently low
to ensure the best time resolution and the maximum detection 
efficiency of the chambers. Taking into account all these conditions
the thresholds were adjusted around 5$-$6 ENC for all chambers.
\begin{table} [h]
\begin{center}
\vspace*{7truemm}
\caption{Characteristics of the readout channels of the different regions.
The average and the r.m.s. of the ENC and of the bias 
experimental distributions are reported in the last 4 columns.}
\label{setting}
\vspace*{8truemm}
\begin{tabular}{|c|c|c|c|c|c|c|c|}
\hline
  &  &\raisebox{-1ex}{$C_{det}$}&\raisebox{-1ex}{Sensitivity}&
\raisebox{-1ex}{$\overline{\mathrm {ENC}}$} &\raisebox{-1ex}{r.m.s. ENC}& 
\raisebox{-1ex}{$\overline{\mathrm {Bias}}$} 
&\raisebox{-1ex}{r.m.s. Bias} \\
\raisebox{3ex}{Region} & \raisebox{3ex}{Readout}&  
\raisebox{1ex}{(pF)}&\raisebox{1ex}{(mV/fC)} 
&\raisebox{1ex}{(fC)}&\raisebox{1ex}{(fC)}&\raisebox{1ex}{(mV)}
&\raisebox{1ex}{(mV)}   
\\ \hline
M1R2 & 	cathode& 58  & 13.3 & 0.41 & 0.08 & 21.1  & 3.9 \\ 
\hline
M1R3 & 	cathode& 51  & 14.0   & 0.35 & 0.09 & 19.7  & 4.3 \\ 
\hline
M1R4 &	anode  & 100 & 10.3 & 0.96 & 0.25 & 23.0  & 2.2 \\ 
\hline
M2R1 &	cathode& 131 & 10.5 & 1.13 & 0.11 & 29.6 & 2.4 \\ 
\hline
M2R1 &	anode  & 75  & 11.3 & 0.71 & 0.12 & 22.6  & 3.1 \\ 
\hline
M2R2 &	cathode& 111 & 11.3 & 0.97 & 0.09 & 29.3 & 2.5 \\ 
\hline
M2R2 &	anode  & 77  & 11.0 & 0.70 & 0.10 & 23.3  & 2.7 \\ 
\hline
M2R3 &	cathode& 89  & 11.9 & 0.76 & 0.09 & 26.6 & 2.5 \\ 
\hline
M2R4 &	anode  & 192 & 8.2  & 1.46 & 0.12 & 28.4 & 2.5 \\ 
\hline
M3R1 &	cathode& 137 & 10.2 & 1.16 & 0.13 & 29.7 & 2.5 \\ 
\hline
M3R1 &	anode  & 81  & 11.0 & 0.60 & 0.07 &  22.3 & 2.9 \\ 
\hline
M3R2 &	cathode& 122 & 10.9 & 1.03 & 0.11 & 29.5 & 2.6 \\ 
\hline
M3R2 &	anode  & 97  & 10.6 & 0.68 & 0.08 & 23.8 & 2.6 \\ 
\hline
M3R3 &	cathode& 91  & 11.7 & 0.80 & 0.09 & 26.8 & 2.4 \\ 
\hline
M3R4 &	anode  & 207 & 7.7  & 1.60 & 0.12 & 28.3 & 2.5 \\ 
\hline
M4R1 &	cathode& 79  & 12.4 & 0.69 & 0.09 & 26.9 & 2.6 \\ 
\hline
M4R2 &	cathode& 127 & 10.4 & 1.12 & 0.10 & 28.5 & 2.4 \\ 
\hline
M4R3 &	cathode& 129 & 10.2 & 1.17 & 0.09 & 28.9 & 2.4 \\ 
\hline
M4R4 &	anode  & 225 & 7.2  & 1.74 & 0.13 & 27.7 & 2.5 \\ 
\hline
M5R1 &	cathode& 82  & 12.3 & 0.72 & 0.09 & 27.1 & 2.7 \\ 
\hline
M5R2 &	cathode& 132 & 9.9  & 1.22 & 0.10 & 28.8 & 2.4 \\ 
\hline
M5R3 &	cathode& 141 & 9.7  & 1.27 & 0.11 & 28.9 & 2.3 \\ 
\hline
M5R4 &	anode  & 245 & 6.7  & 1.89 & 0.13 & 26.9 & 2.2 \\ 
\hline
\end{tabular}
\end{center}
\end{table}
\vspace*{2truemm}

\subsection{Malfunctionning channels and ageing effects}
\label{expres2}
\vspace*{7truemm}

\begin{figure}[t]
\begin{center}
\includegraphics[width=.85\textwidth]{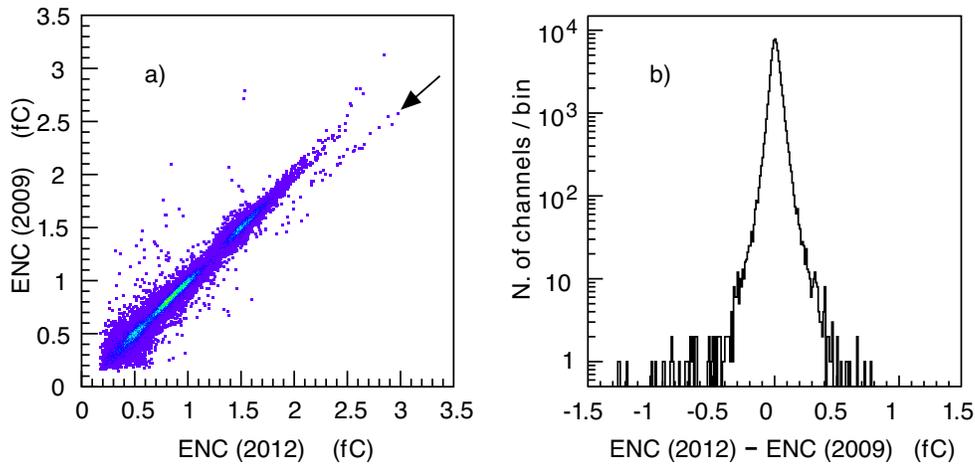}             
\end{center}
\vspace*{-6truemm}
\caption {
a): Scatter plot of the ENC values measured on all the muon detector in 
the years 2009 and 2012. The channels which are far from the plot bisector 
were checked and repaired or replaced.
Most of the points in the series indicated by an arrow correspond to 
CARIOCAs mounted on the same boards which were replaced. 
b): Distribution of the difference between the ENC 
measured in 2012 and 2009.
}
\label{scatter_09-12}             
\end{figure}

The method described allows to detect breakdown of some front-end or 
any possible ageing effects, by checking the stability of the ENC 
for all the readout channels. 
In figure~\ref{scatter_09-12}a the ENC measured in the year 2012 is
compared with that measured in 2009. Most of the channels are
stable, and only few tens, out of 122k, show a significant change
of their ENC value and need to be individually checked, repaired
or replaced.
In particular a series of channels (indicated by an arrow in 
figure~\ref{scatter_09-12}a) appear to be aligned out of the
scatter plot bisector which suggests a possible failure in the correct 
threshold setting by the DACs. Most of these channels belonged to the same 
boards, which were replaced.
The distribution of the difference between 
the ENC measured in the years 2009 and 2012 is reported in 
figure~\ref{scatter_09-12}b. The large peak around zero shows that
most of the frontends are quite stable over many years and no ageing
effects were observed up to 2~fb$^{-1}$, which corresponds to about
one tenth of the total integrated luminosity expected in 10 years 
operation.

%
\begin{figure}[h]
\begin{center}
\vspace*{4truemm}
\includegraphics[width=.85\textwidth]{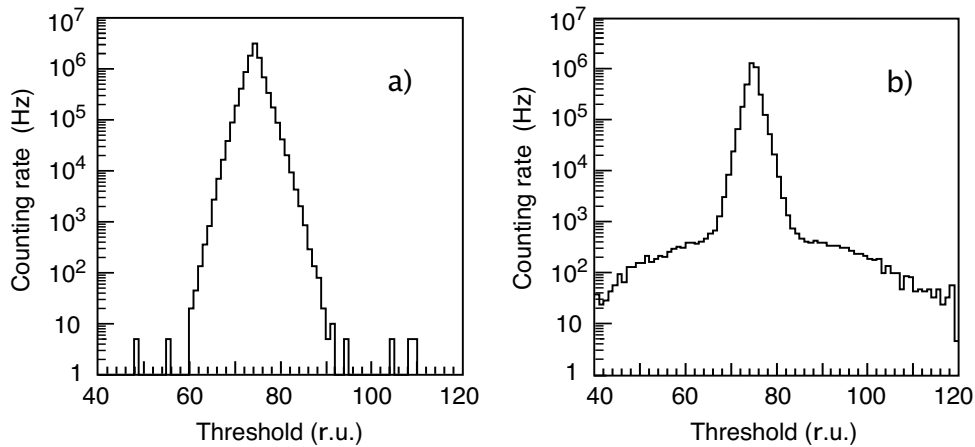}
\end{center}
\vspace*{-4truemm}
\caption{
Experimental threshold scanning for: a) a properly behaving front-end channel; 
b) a malfunctioning channel with an abnormal noise.
}
\label{noisy}
\end{figure}

Another signal of malfunctioning
of a channel readout is an abnormal shape of the threshold scan.
A typical example in shown in figure~\ref{noisy}. 

\section{Conclusions}
A new method has been introduced which allows to monitor the correct 
operation of the front-end channels of the LHCb muon detector 
and to set the optimal discriminator thresholds.
The method consists in measuring the noise rate as a function of 
the discriminator thresholds. The resulting threshold scan distribution 
is fitted to a tail of a Gaussian as suggested by Rice theory
and confirmed by a Monte Carlo simulation. The fitting procedure is
greatly facilitated by the knowledge of the counting rate at zero 
threshold, which cannot be measured but can be calculated from the 
amplifier bandwidth using the Rice's formulas. 
For each threshold scan of a readout channel, the r.m.s. of the fitted 
Gaussian tail turns out to 
be equal to the equivalent noise charge (ENC) of that channel.
The measurement of the ENC of all the channels allows to check for
a possible malfunction or breakdown and to set the working values
of the thresholds which satisfy the requirements of a low
counting rate due to electronic noise and a high detection efficiency and 
time resolution of the chambers.



\begin{thebibliography}{99}
%
\bibitem{jinst}
A. Augusto Alves Jr. et al., \emph{The LHCb Detector at the LHC}, 
\jinst{3}{2008}{S08005}.
%
\bibitem{muonpaper}
A. Augusto Alves Jr. et al., \emph{Performance of the LHCb muon system},
arXiv:1211.1346, submitted to JINST.
%
\bibitem{carioca}
W.Bonivento et al., \emph{Development of the CARIOCA 
front-end chip for the LHCb muon detector},  
\emph{Nucl. Instrum. Meth.} {\bf A 491} (2002) 233.
%
\bibitem{gain}
E. Dan\'e, et al., \emph{Detailed study of the gain of the MWPCs for the 
LHCb muon system}, 
\emph{Nucl. Instrum. Meth.} {\bf A 572} (2007) 682.
%
\bibitem{burkhard}
L. Gruber, W.Riegler and B.Schmidt, \emph{Time 
resolution limits of the MWPCs for the LHCb muon system}, 
\emph{Nucl. Instrum. Meth.} {\bf A 632} (2011) 69.
%
\bibitem{anatoli}
A. P. Kashchuk and O. V. Levitskaya, \emph{From noise to signal- a new 
approach to LHCb muon optimization},
\emph{LHCb note, LHCb-PUB-2009-018} (2009),
(http://cdsweb.cern.ch/record/1209624/files/LHCb-PUB-2009-018.pdf)
and references therein quoted.
%
\bibitem{anatoli2}
A. P. Kashchuk, \emph{Application of the Rice theory for reconstructing 
the noise distributions in nuclear electronics},
\emph{Instruments and Experimental Techniques} {\bf 55} (2012) 440, 
engl. transl. of \emph{Pribory i Tekhnika Eksperimenta} {$N^o$}~4 
(2012) 26.
%
\bibitem{lukas}
L. Gruber, \emph{Optimization of the operating parameters of
the LHCb muon system}, CERN-THESIS-2010-037 (2010).
%
\bibitem{rice}
S. O. Rice, \emph{Mathematical analysis of random noise}, 
\emph{Bell System Technical Journal} {\bf 23} (1944) 282 
and {\bf 24} (1945) 46.
%
\bibitem{kac}
M. Kac, \emph{On the Distribution of Values of Trigonometric Sums with 
Linearly Independent Frequencies}, \emph{Amer. Jour. Math.} {\bf 65} 
(1943) 609. 
%
\bibitem{ivanov}
V. A. Ivanov, \emph{On the Average Number of Crosssings of a Level by 
Sample Functions of a Stochastic Process},
\emph{Theory of Probability and Its Applications}, engl. transl. of
\emph{Teor. Veroyatnost. i Primen.} {\bf 5} (1960) 319. 
%
\bibitem{longuet}
M. S. Longuet-Higgins, \emph{The Distribution of Intervals between Zeros 
of a Stationary Random Function},
\emph{Phil. Trans. Royal Society of London; Series A, Math. and Phys. 
Sciences} {\bf 254} (1962) 557 (http://www.jstor.org/stable/73149),
and references therein quoted.
%
\bibitem{anghinolfi}
F. Anghinolfi, \emph{Front-end Electronics}, \emph{Course on Detectors and 
Electronics for High Energy Physics, Astrophysics and Space Applications},
INFN Laboratori Nazionali di Legnaro, 26-30 March 2007.

%
\end{thebibliography}
\end{document}